\documentclass[12pt]{iopart}

\usepackage{enumitem}
\usepackage{amssymb}
\usepackage{amsfonts}
\usepackage{braket}
\usepackage{dsfont}
\usepackage{MnSymbol} 
\newcommand{\imag}{\mathrm{i}}
\usepackage{color}
\usepackage[dvipsnames]{xcolor}
\usepackage{dirtytalk}
\usepackage{tikz}
\usepackage{bbm}

\newcommand{\new}[1]{\textcolor{black}{#1}}

\usepackage{graphicx, import}
\usepackage{soul}
\usepackage[utf8]{inputenc}
\usepackage[export]{adjustbox}
\usepackage[caption=false]{subfig}
\usepackage{parskip} 
\usepackage{hyperref}
\usepackage{ulem}

\usepackage{xspace}

\begin{document}
\title[Recompilation-enhanced simulation of electron-phonon dynamics]{Recompilation-enhanced simulation of electron-phonon dynamics on IBM Quantum computers}

\author{Benjamin Jaderberg$^{1}$, Alexander Eisfeld$^{2}$, Dieter Jaksch$^{1,3}$, and Sarah Mostame$^{4}$}
\address{$^1$Clarendon Laboratory, University of Oxford, Parks Road, Oxford OX1 3PU, United Kingdom \\
$^2$Max Planck Institute for the Physics of Complex Systems, Nöthnitzer Straße 38, 01187 Dresden, Germany\\
$^3$Institut für Laserphysik, Universität Hamburg, 22761 Hamburg, Germany \\
$^4$IBM Quantum, IBM T.J. Watson Research Center, Yorktown Heights, NY 10598, USA} 
\date{\today}

\begin{abstract}
Simulating quantum systems is believed to be one of the first applications for which quantum computers may demonstrate a useful advantage. For many problems in physics, we are interested in studying the evolution of the electron-phonon Hamiltonian, for which efficient digital quantum computing schemes exist. Yet to date, no accurate simulation of this system has been produced on real quantum hardware. In this work, we consider the absolute resource cost for gate-based quantum simulation of small electron-phonon systems as dictated by the number of Trotter steps and bosonic energy levels necessary for the convergence of dynamics. We then apply these findings to perform experiments on IBM quantum hardware for both weak and strong electron-phonon coupling. Despite significant device noise, through the use of approximate circuit recompilation we obtain electron-phonon dynamics on current quantum computers comparable to exact diagonalisation. Our results represent a significant step in utilising near term quantum computers for simulation of quantum dynamics and highlight the novelty of approximate circuit recompilation as \new{a tool for reducing noise}.
\end{abstract}
    
\maketitle

\section{Introduction} \label{sec:introduction}

Since its inception, the number of potential applications for quantum computers has grown tremendously, including at present, computational chemistry and energy minimisation \cite{Kandala2017VQE,peruzzo2014variational,Cao2019review,RevModPhys.92.015003}, pattern finding \cite{biamonte2017quantum,havlivcek2019supervised} and optimisation \cite{farhi2014quantum,moll2018quantum}. Yet the number of algorithms with provable speedup over their classical counterparts is small \cite{shor1999polynomial, arute2019quantum, grover1996fast} and further still when discounting those which aren't useful \cite{deutsch1992rapid}. Of the applications remaining, simulating the evolution of quantum systems is perhaps the most promising, with its exponential advantage \cite{lloyd1996universal} in memory and time over classical simulation and broad range of use cases. For quantum computers built on two-level qubits, the simulation of purely fermionic systems has been explored widely due to their naturally equivalent degrees of freedom. However, it is the inclusion of vibrational modes that gives rise to specific important phenomena such as phonon-mediated superconductivity \cite{marsiglio2008electron}. Thus, the ability to accurately simulate systems with both electronic and vibrational modes on a quantum computer could result in significant technological advances.

One approach to simulating electron-phonon systems is through the use of analogue quantum devices \cite{mostame2017emulation,PhysRevLett.114.123005,PhysRevB.88.224502,PhysRevLett.109.250501,PhysRevA.97.052321,PhysRevResearch.2.023133,Anton2018FMO, johnson2011impurity,bruderer2007polaron,bruderer2008transport}. This can be achieved by representing the electronic components of the system with qubits, whilst the phonon modes are represented by a physical system with the desired vibrational interactions, such as inductor–resistor–capacitor oscillators \cite{mostame2012quantum}, standing light waves \cite{stojanovic2012quantum} or Rydberg atoms \cite{hague2014cold}, depending on the type of qubit. Whilst analog quantum simulators continue to produce impressive results and address interesting problems in physics and chemistry, their application is nevertheless limited by their inflexibility and non-universality. By contrast, a digital quantum computing approach could modify the coupling strength, topology, system size and even introduce additional interactions with no cost.

Implementing d-level systems (e.g., phonons and other bosons) on a quantum computer can be achieved in a number of ways, including  a hybrid quantum-classical approach discussed in \cite{magann2021digital} or both first \cite{ollitrault2020nonadiabatic,macridin2018digital,PhysRevLett.121.110504} and second quantisation mappings \cite{somma2003quantum} of the wavefunction. In this work we focus on the latter, for which schemes were recently demonstrated requiring an efficient number of qubits and circuit operations \cite{sawaya2020resource}. Such considerations are important if simulations are going to be run on near-term quantum computers, which are limited in both their size and number of operations that can applied before the qubits decohere. Nevertheless, despite efficient scaling, accurate evolution of the electron-phonon Hamiltonian has not been demonstrated on real quantum hardware to date, owing to the large number of gates required for even one Trotter step. 

In this work we study the dynamics of electron-phonon systems through experiments on the Qiskit~\cite{qiskit} statevector simulator (SV) and real quantum hardware and consider the limits required to obtain accurate results in each case. In section \ref{sec:method} we introduce the electron-phonon Hamiltonian and detail the circuit implementation of its time propagator, including initialisation and measurement of the \new{electronic excitation} population. In section \ref{sec:results} we present our results, starting in section \ref{subsec:ED_comparison_results} with establishing the qubit and Trotter step requirements for small systems to obtain accurate dynamics. Following this, in section \ref{subsec:real_device_results} we apply these findings to simulate systems with up to 3 electronic sites on real quantum computers. Through the use of incremental structural learning (ISL) \cite{jaderberg2020minimum}, an approximate circuit recompilation method, we maximise the performance of current quantum hardware to achieve results that match exact diagonalisation in the strong and weak coupling regimes. Finally, in section \ref{sec:conclusion} we summarise our findings.

\section{Method}\label{sec:method}

\subsection{Model}

The electron-phonon model is used to describe many systems with electronic and vibrational degrees of freedom, such as molecular aggregates \cite{hestand2018expanded} including natural and artificial light harvesting systems \cite{brixner2017exciton}. In this model we consider $N$ sites, where each site consists of an electronic two-level system. The electronic part of the Hamiltonian describing this model is given by
\begin{equation}\label{eqn:Hel}
   H_{\mathrm{el}} = \sum^{N}_{i=1} \tilde{\epsilon}_i|i\rangle \langle i| + \sum_{j=1}^N \sum^{N}_{i<j}V_{ij}(|i\rangle \langle j| + |j\rangle \langle i|)
\end{equation}
where $\tilde{\epsilon}_i$ is the electronic transition energy for site $i$ and $V_{ij}$ is the dipole-dipole coupling between sites $i$ and $j$. Here we focus on the case of a single electronic excitation, such that $\ket{i}=\ket{g_{1}\cdots e_{i}\cdots g_{N}}$ denotes the state where all sites are in the electronic ground state, except site $i$ which is populated by an excitation. In this framework, we can also consider $V_{ij}$ as the hopping amplitude of the excitation between sites.
The vibrational modes of the system can be described under the harmonic approximation by a finite set of quantum harmonic oscillators (QHOs) \cite{mcardle2019digital, macridin2018digital}. In this case, the phonon Hamiltonian can be written as
\begin{equation}\label{eqn:Hph}
   H_{\mathrm{ph}} =\sum_{i=1}\sum_{l}\hbar\omega_{il}({\hat{a}^{\dagger}_{il}}{\hat{a}_{il}} + 1/2)
\end{equation}
where $\hat{a}^{\dagger}_{il}$ ($\hat{a}_{il}$) is the bosonic creation (annihilation) operator for the vibrational excitation of the $l^{\mathrm{th}}$ phonon mode of site $i$ and $\omega_{il}$ is the frequency of oscillation.

\begin{figure}
	\centering
	\includegraphics[width=8cm]{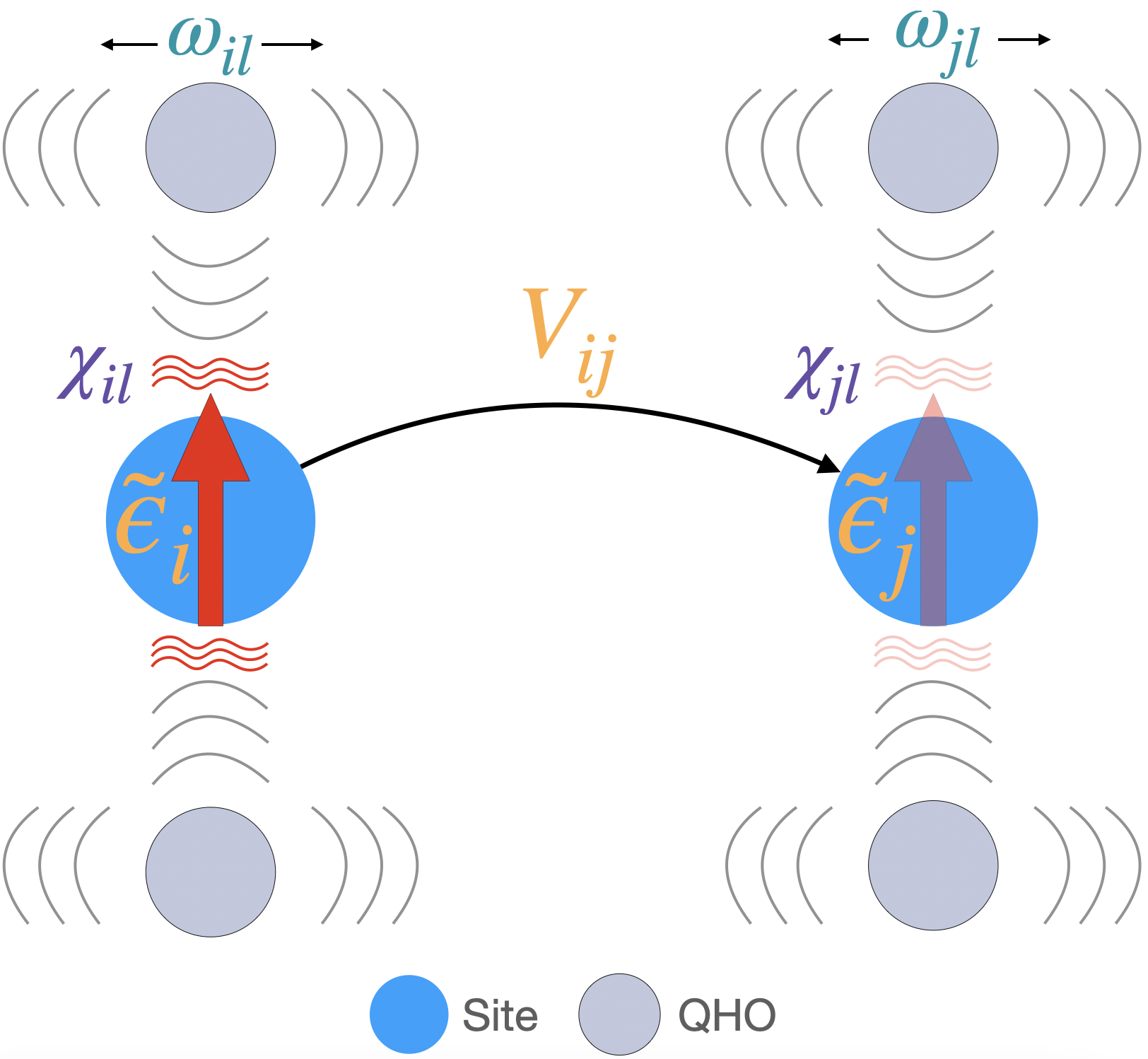}
	\caption{Diagram of the model considered in this work. A lattice site $i$ can be occupied by an electronic excitation with transition energy $\tilde{\epsilon}_{i}$, which can hop to connected sites with transition amplitude $V_{ij}$. Each site is also connected to $l$ phonon modes, approximated by quantum harmonic oscillators with frequency $\omega_{il}$, which interact with the excitation via the electron-phonon coupling constant $\chi_{il}$.}
	\label{fig:system_diagram}
\end{figure}

The interactions between the electronic excitations and phonons is given by
\begin{equation}\label{eqn:Hep}
   H_{\mathrm{ep}} = \sum^{N}_{i=1} |i\rangle \langle i| \left[\sum_{l}\chi_{il}({\hat{a}^{\dagger}_{il}} + {\hat{a}_{il}}) \right]
\end{equation}
where $\chi_{il}$ is the interaction strength between electronic site $i$ and it's $l^\mathrm{th}$ phonon mode and has dimension of energy. Finally, the total electron-phonon Hamiltonian can be expressed as the sum of these three components
\begin{equation}\label{eqn:total_hamiltonian}
    H = H_{\mathrm{el}} + H_{\mathrm{ph}} + H_{\mathrm{ep}}.
\end{equation}
The interactions described in the electron-phonon Hamiltonian and their corresponding parameters can be seen in Figure \ref{fig:system_diagram}. We note that for this Hamiltonian each QHO is local to a single site, which models the internal vibrational modes of molecular aggregates, and as such extensions would be required to simulate systems with shared modes. \new{ We further note that within this model there is no restriction on the number of vibrational quanta. However, the number of electronic excitations is conserved, as is evident from the Hamiltonian in Eq. (\ref{eqn:Hel}). In real molecules the electronic excitation would decay radiatively or non-radiatively to the ground state, yet this decay is usually much slower than the transfer of excitation between sites.  Interaction with laser pulses could  also increase the number of electronic excitations. For more information on the model we refer readers to Refs.~\cite{MaKu11__,AmVaGr00__}.}

\subsection{Circuit implementation}\label{subsec:circuit_implementation}

To study the dynamics of electron-phonon systems we need to simulate the action of the time evolution operator ${\hat{U}(\tau)} = \exp({-\imag\hat{H}\tau/\hbar})$ on some initial state $|\psi_\mathrm{init}\rangle$. This can be implemented on a quantum computer by considering the logic gates that implement the exponential of each term in Eq. (\ref{eqn:total_hamiltonian}), the products of which approximates the time evolution operator via a first order Suzuki-Trotter decomposition~\cite{suzuki1976generalized}
\begin{equation}\label{eqn:u_tau}
    {\hat{U}(\tau)} = \left(e^{\frac{-\imag H_{el}\tau}{\hbar\eta}} e^{\frac{-\imag H_{ph}\tau}{\hbar\eta}} e^{\frac{-\imag H_{ep}\tau}{\hbar\eta}}\right)^\eta
\end{equation}
to within error $O(\delta^2)$ where $\delta = \frac{\tau}{\eta}$ and $\eta$ is the number of steps the evolved time is discretised into, often called the Trotter steps.

\begin{figure}
	\centering
	\includegraphics[width=8cm]{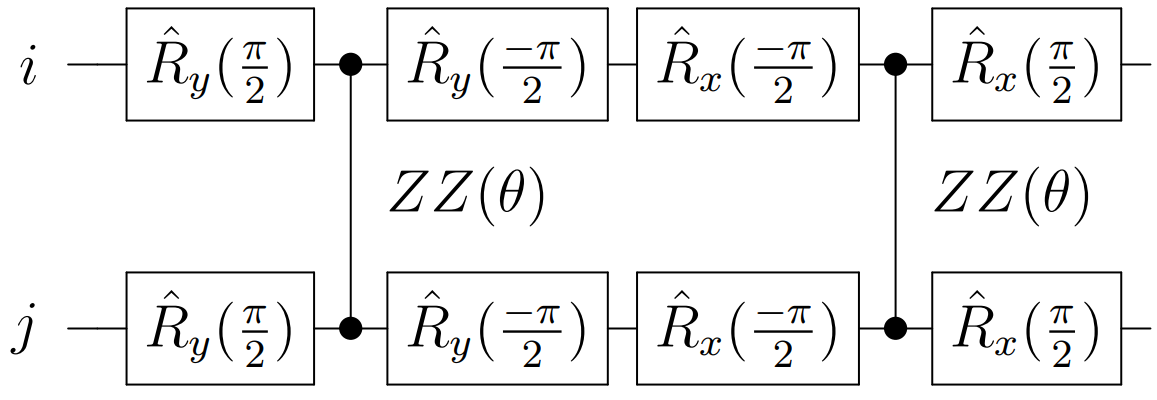}
	\caption{Circuit diagram implementing the operator $\hat{XY} = \exp{\frac{-\imag\theta}{2}(\hat{\sigma}^x_i \hat{\sigma}^x_j + \hat{\sigma}^y_i \hat{\sigma}^y_j)}$.}
	\label{fig:hopping_circuit}
\end{figure}

The electronic part of the Hamiltonian $H_{\mathrm{el}}$ can be implemented on a quantum computer by rewriting it in terms of Pauli matrices. Substituting the site operators for qubit operators through the relations found in \ref{app:qubit_to_pauli} we obtain
\begin{equation}\label{eqn:h_el_paulis}
    H_{el} = \frac{1}{2} \sum^{N}_{i=1} \tilde{\epsilon}_i \; \hat{\sigma}^z_i + \frac{1}{2} V_{ij}(\hat{\sigma}^x_i \hat{\sigma}^x_j + \hat{\sigma}^y_i \hat{\sigma}^y_j),
\end{equation}
where $\hat{\sigma}_{i}^{\alpha}$ is the Pauli operator $\alpha \in \set{x,y,z}$ acting on qubit $i$. Exponentiating Eq. (\ref{eqn:h_el_paulis}), each term of the first summation can be implemented by a single qubit rotation $\hat{R}^z_i(\theta) = \exp{-\imag\theta\hat{\sigma}^z_i/2}$, where substituting the general rotation parameter $\theta = \tilde{\epsilon}_i\tau/\eta$ gives us the desired evolution for one Trotter step. For the second summation in Eq. (\ref{eqn:h_el_paulis}), the exponential of each term can be implemented by the circuit shown in Figure. \ref{fig:hopping_circuit} \cite{kreula2016few}. Here simulating one Trotter step of evolution requires the $ZZ$ rotation $\theta=V_{ij}\tau/\eta$.

To implement $H_{ph}$ and $H_{ep}$, we first require a representation of the QHO in the qubit basis. This is achieved using the standard binary mapping, a method which although established in previous works \cite{veis2016quantum, mcardle2019digital, sawaya2019quantum, sawaya2020near}, we briefly describe here for completeness. Considering a QHO in the second quantisation formalism, we represent each of its $d$ energy levels by binary strings (e.g., $|0\rangle, |1\rangle, |2\rangle, |3\rangle \longrightarrow |00\rangle, |01\rangle, |10\rangle, |11\rangle$), which can subsequently be encoded into the computational basis states of $n_x = \log_{2}(d)$ qubits. Since the QHO has infinite bosonic degrees of freedom, any value of $d$ will introduce truncation errors. Therefore, it is important to find the number of levels required for the dynamics to converge, as we do later.

In the second quantisation, any operator acting across the QHO energy levels can be expressed in terms of the ladder operators $\hat{a}^\dagger$, $\hat{a}$. In turn, these can be written as binary operators and subsequently converted to Pauli matrices, as shown in \ref{app:ladder_to_qubit}. Overall, using this method allows the transformation of either ladder operator to $O(d^2)$ Pauli strings \cite{mcardle2019digital}, which we define as $\hat{P} = \bigotimes_{i=1}^{n_x} \hat{\sigma}^\alpha_i$. For our time evolution algorithm, we implement the exponential $\exp(-\imag \theta \hat{P})$ in a quantum circuit using the well-known CNOT staircase method \cite{whitfield2011simulation, sawaya2020resource, yordanov2020efficient}.

Using the qubit representations of the QHO described above, we can decompose the time evolution of the remaining Hamiltonians $H_{ph}$ and $H_{ep}$ into quantum circuits. For $H_{ph}$, the bosonic number operator ${\hat{a}^{\dagger}_{il}}{\hat{a}_{il}}$ is a diagonal evenly-spaced operator which has the known decomposition \cite{sawaya2020resource}
\begin{equation}
    {\hat{a}^{\dagger}_{il}}{\hat{a}_{il}} = \sum_{k=0}^{n_x}2^k\hat{\sigma}^z_{k+1} = 2^0\hat{\sigma}^z_{1} + 2^1\hat{\sigma}^z_{2} + ...
\end{equation}
where the index $k$ spans specifically the qubit register representing the $l^{\mathrm{th}}$ phonon mode of site $i$. Thus
\begin{subequations}
    \begin{align}
        \exp(-\imag\theta {\hat{a}^{\dagger}_{il}}{\hat{a}_{il}}) &= \prod_{k=0}^{n_x}\exp(-\imag\frac{\theta}{2} 2^{k+1} \hat{\sigma}^z_{k+1}) \\
        &= \prod_{k=1}^{n_x}\hat{R}_{z}^{k}(-\imag\theta2^{k}),
    \end{align}
\end{subequations}
implements the desired Trotterised phonon evolution for each term in Eq. (\ref{eqn:Hph}) given $\theta = \hbar\omega_{il}\tau/\eta$.

Finally, the time evolution circuit for $H_{ep}$ is generated first by following the decomposition of $\hat{a}^\dagger_{il} + \hat{a}_{il}$ in Eq. (\ref{eqn:expanding_adagger_plus_a}) through (\ref{eqn:pauli_string_example}), continuing to generate Pauli strings $\hat{P}_{il}$ for all the binary operators. Combining these with the prepended site operator, the terms $\hat{\sigma}^z_j \bigotimes \hat{P}_{il}$ build up the Trotter step circuit $\exp(-\imag H_{ep}) = \prod_{i}\prod_{l}\exp(-\imag\theta \hat{\sigma}^z_i \bigotimes \hat{P}_{il})$, where each term in the product is implemented using the previously cited CNOT staircase method. This implements the Trotterised evolution of $H_{ep}$ given $\theta = \chi_{il}\tau/\eta$.

\subsection{Calculation of population}

\begin{figure}
	\centering
	\includegraphics[width=8cm]{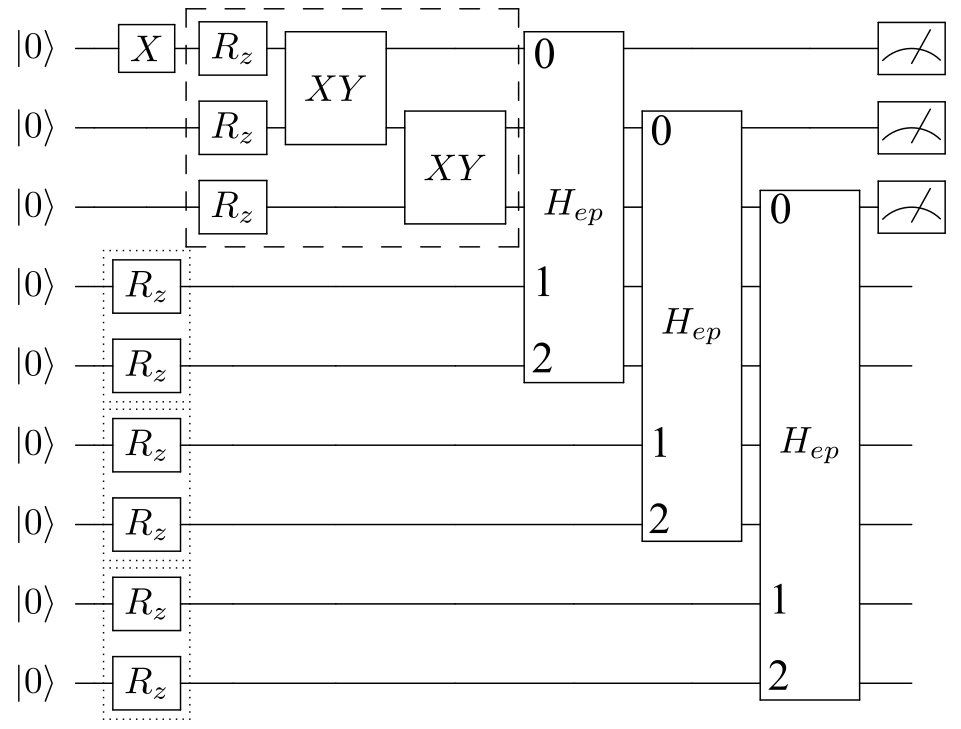}
	\caption{Circuit diagram showing state initialisation and a single time evolution step. First, one of the site qubits are flipped to initialise the system with a single excitation, in this case site 0. The dashed box shows the application of $H_{el}$, with phase rotation gates to encode the electronic transition energies and $\hat{XY}$ gates encoding the \new{electronic excitation} hopping. Note that in this case, the sites are not fully connected and instead represent a model with a chain topology. The dotted boxes shows the application of $H_{ph}$ to each oscillator, each of which consist of a $n_x=2$ qubit register for a total of $d=4$ energy levels. Afterwards, the interaction $H_{ep}$ between each site and its corresponding phonon mode is applied, the decomposition of which is too large to show. \new{In the diagram, $H_{ep}$ includes explicit internal indices of the site (0) and the two qubits that make up the phonon register (1, 2) to indicate that the other qubit wires that pass underneath are not acted on.} Finally, the site qubits are measured to obtain observables such as \new{population}.}
	\label{fig:ts_circuit}
\end{figure}

In section \ref{subsec:circuit_implementation} we outlined the circuits required to implement one Trotter step of the electron-phonon Hamiltonian. Simulating the time dynamics of a system which evolves under this Hamiltonian requires the application of multiple Trotter steps on an initial state $|\psi_\mathrm{init}\rangle$, which in this work we consider to be the electronic and vibrational ground state apart from a single electronic excitation. This is achieved by initialising all of the qubits in the $|0\rangle$ state and then performing a bit flip on the one site where the excitation is localised to, requiring only 1 gate in total. We note that this is an advantage compared to first quantisation schemes \cite{macridin2018digital}, in which the vibrational ground state requires the preparation of a Gaussian across the amplitudes of the phonon qubits. The overall circuit for state initialisation followed by a single Trotter step can be seen in Figure. \ref{fig:ts_circuit}.

Once the desired time evolution has been reached by repeated Trotter steps, the qubits are measured to retrieve the desired time-dependent observable under study. Here we use a single measurement of each site in the $\hat{\sigma}^z$ basis, which results in the qubits collapsing to the binary $|0\rangle$ or $|1\rangle$ state depending on the absence or presence of the excitation respectively. Through repeated sampling of the circuit we obtain the average \new{electronic excitation} population on each site at the given time. \new{From here we refer to this simply as the population.}

\section{Results}\label{sec:results}

In this section we present experiments simulating the time evolution of electron-phonon systems on both classically emulated and real quantum computers.

\subsection{Simulator results with comparison to ED}\label{subsec:ED_comparison_results}

We first apply our algorithm to small systems of only two electronic sites and one phonon mode per site. This serves two purposes. Foremost, this allows us to verify our gate-based quantum computing implementation of the time evolution propagator and compare the results obtained to exact diagonalisation (ED) methods. Secondly, this allows us to check the convergence of dynamics with respect to errors caused by our mapping of the Hamiltonian into a qubit basis, namely Trotterisation error and truncation from representing infinite bosonic degrees of freedom with a finite number of qubits. In this section, for simplicity we only consider the case whereby site dependent parameters are equal (e.g., $V_{ij} = V_{kl} = V$), and subsequently set $\tilde{\epsilon}= 0$ since it only contributes a global phase. Furthermore, we set our unit of energy to be $\hbar\omega$. Proof of the algorithms performance for larger simulations with non-uniform parameters can be found in \ref{app:large_results}.

\subsubsection{Weak electronic coupling}

\begin{figure}
	\centering
	\includegraphics[width=\linewidth]{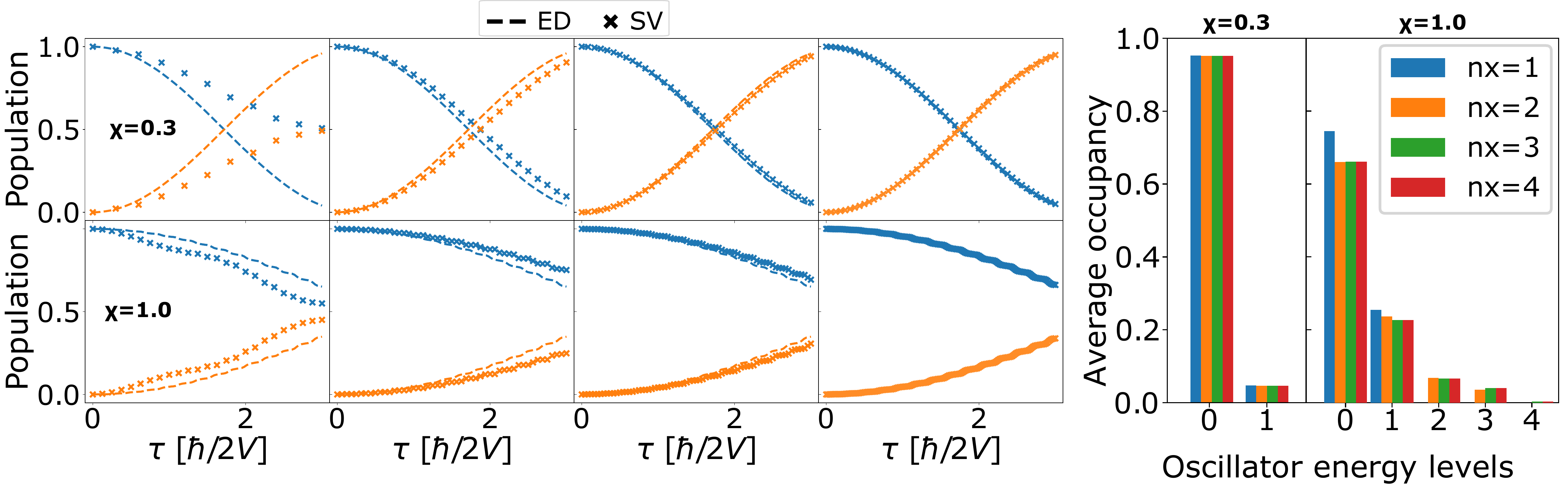}
	\caption{Left: \new{population} over time at electronic sites 0 (blue) and 1 (orange), as calculated by ED (dashed line) or quantum circuits run on the Qiskit Aer statevector simulator (SV, crossed marker). The Hamiltonian is simulated with \new{system parameters $N=2$,} $V=0.05\hbar\omega$, $\chi=0.3\hbar\omega$ (upper panels), $\chi=1.0\hbar\omega$ (lower panels) and $n_{x}=1$. Moving across the panels left to right, for the quantum circuit approach, time evolution is discretised by 10, 24, 36 and 48 Trotter steps for $\chi=0.3\hbar\omega$ and 24, 48, 72 and 144 Trotter steps for $\chi=1.0\hbar\omega$. Right: average energy level occupation of the oscillator coupled to site 0 during time evolution. The blue bar corresponds to the experiment shown on the left, whilst the other bars represent subsequent experiments with increasing $n_x$. These experiments were run with the number of Trotter steps required for convergence (e.g., $\eta = 48$ for $\chi=0.3\hbar\omega$ and $\eta = 144$ for $\chi=1.0\hbar\omega$).}
	 \label{fig:v=0.05}
\end{figure}

Figure \ref{fig:v=0.05} shows the population for the case of weak electronic coupling $V=0.05\hbar\omega$ with non-zero electron-phonon coupling. Starting with $\chi=0.3\hbar\omega$ on the upper panels, the ED results show that overall, the dynamics look very similar to the purely electronic case (see \ref{app:chi_0_results}) except that the crossing where the populations are equal occurs slightly later. This is caused by excitation of the vibrational modes of the system, leading to an effective re-scaling of the electronic coupling \cite{roden2009electronic}. However, the results as calculated using quantum circuits on the SV simulator are different to the purely electronic case in terms of convergence with number of Trotter steps. With the electron-phonon interactions now switched on, the non-commuting terms of the Hamiltonian lead to an error in the observed population when using a small number of Trotter steps. As we increase the number of Trotter steps, we see that this error reduces, reaching convergence with the ED solution at $\eta=48$.

Next we consider the strong electron-phonon coupling case $\chi=1.0\hbar\omega$, shown on the lower panels of Figure \ref{fig:v=0.05}. Here the dynamics evolve differently to $\chi=0.3\hbar\omega$. Firstly, the time taken for the excitation to transfer between electronic sites is further extended, to the point where the crossing doesn't occur within the simulated time. Furthermore, we also observe a second mode of population transfer, visible as rapid oscillations. Looking across at the Trotter step convergence, we find that $\eta=144$ Trotter steps are required to produce these more complicated dynamics with accuracy equal to ED.

After finding the number of Trotter steps required to accurately simulate the dynamics for $V=0.05\hbar\omega$, we next consider convergence with respect to the truncation of the QHO. In the furthest right graph of Figure \ref{fig:v=0.05}, the blue bar shows the average occupation of each energy level of the oscillator coupled to site 0 during the converged experiments on the left. This is then repeated with increasing numbers of QHO energy levels $d=2^{n_x}$ up to $n_x = 4$. Looking at the results of all the bars, for $\chi=0.3\hbar\omega$ we find that only the ground state $d=0$ and first excited state $d=1$ are occupied to any observable level. This implies that simulation of this system is indeed converged with only $n_x=1$ qubit per phonon mode. For strong electron-phonon coupling however, the truncation when choosing $n_x=1$ is clear, since for larger values of $n_x$ there is a non-zero occupation of the $d=2$ and $d=3$ excited states. Furthermore, whilst the $d=4$ level has a very small average occupation of $0.003$ and is barely visible, we can conclude that its inclusion is necessary for convergence of the dynamics. This can be seen in the $n_x=2$ case, whereby not being able to access the $d=4$ excited state leads to an overestimation of the occupancy at $d=1$ and underestimation at $d=3$. Thus, for a two-site $V=0.05\hbar\omega, \chi=1.0\hbar\omega$ system, each phonon mode needs to be represented by at least $n_x=3$ qubits for full convergence. 

\subsubsection{Strong electronic coupling}

\begin{figure}
	\centering
	\includegraphics[width=\linewidth]{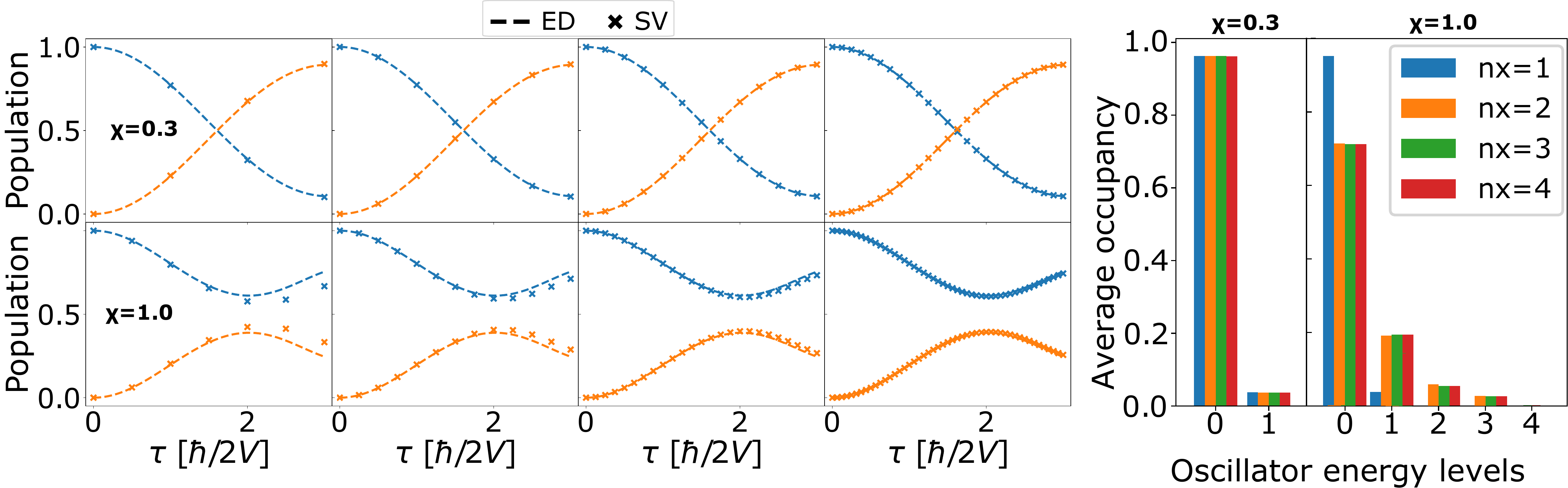}
	\caption{Left: \new{population} over time at electronic sites 0 (blue) and 1 (orange), as calculated by ED (dashed line) or quantum circuits run on the statevector simulator (crossed marker). The Hamiltonian is simulated with \new{system parameters $N=2$,} $V=1.0\hbar\omega$, $\chi=0.3\hbar\omega$ (upper panels), $\chi=1.0\hbar\omega$ (lower panels) and $n_{x}=1$ with increasing Trotter steps $\eta=$ 3, 6, 12 and 24 Trotter steps for $\chi=0.3\hbar\omega$ and $\eta=$ 6, 12, 24 and 48 for $\chi=1.0\hbar\omega$. Right: average energy level occupation of the oscillator coupled to site 0 during time evolution.}
	 \label{fig:v=1.0}
\end{figure}

Next we consider our two-site system in the strong electronic coupling regime $V=1.0\hbar\omega$, repeating the experiments to determine convergence with respect to $\eta$ and $n_x$. These results can be seen in Figure \ref{fig:v=1.0}. Notably, although the larger value of $V$ leads to the excitation hopping between sites at a much faster rate, this is factored out by plotting time in units of $\hbar/2V$. With this accounted for, looking at the results for $\chi=0.3\hbar\omega$ on the upper panels, we see that the population looks quite similar to its equivalent in Figure \ref{fig:v=0.05}, albeit with a slight damping of the oscillation. However, there is a large difference when comparing convergence with respect to Trotter steps, which occurs here for as few as $\eta=6$ as shown in the second panel. Unlike the other regimes studied in this work, such few Trotter steps are required for an accurate solution because it is the closest to the purely electronic case based on the ratio V/$\chi$. Thus, whilst this regime may be the first that can be simulated accurately on near term quantum computers, due to its low $\eta$ requirements, it also has one of the most trivial dynamics. This is further demonstrated by the average oscillator energy level occupancy, as shown in the rightmost graph. Here we see the state of the oscillator can be represented without truncation by only $n_x=1$ qubit, with only the ground and first excited state being occupied.

Finally we examine the case of both strong electronic coupling $V=1.0\hbar\omega$ and strong electron-phonon coupling $\chi=1.0\hbar\omega$, shown in the lower panels of Figure \ref{fig:v=1.0}. Here, unlike the weak regime, the populations do not cross and we instead see an inflection at $\tau=2$. Simulation over a longer time would be required to see whether this is in fact a secondary higher frequency mode of oscillation, or is a reflection of the primary excitation wave. Looking across the panels, we see that $\eta=48$ Trotter steps are required for convergence. Considering the average QHO energy level occupancy, the rightmost graph demonstrates again that in the strong electron-phonon regime the oscillators are excited to higher energy levels than the weak regime. Similar to the $V=0.05\hbar\omega$ case, whilst the scarcely visible $d=4$ occupation may suggest that $n_x=2$ is sufficient, the average occupancy across all levels only converges for $n_x=3$ qubits.

\subsection{Real device results}\label{subsec:real_device_results}

Up until this point, all results have been obtained by executing the quantum circuits that make up the time evolution operator on a classically emulated quantum computer. This was done to factor out the noise generated by real quantum computers, which for current generation devices is prohibitively large to accurately simulate the time evolution circuits required. However, in this section we show that with approximate circuit recompilation, it is in fact possible to obtain accurate results which capture the dynamics of the electron-phonon Hamiltonian on a real quantum computer.

Given a unitary $\hat{U}$, quantum circuit compilation describes the process of finding an implementation of $\hat{U}$ on a quantum computer using a series of $m$ gate operations. On the other hand, quantum circuit \textit{re}compilation is the process of trying to improve an existing circuit by replacing it with a circuit that generates the same unitary but takes less time to run. This is important for current quantum computers, whose short-lived qubits rapidly undergo decoherence, creating noise, and as such quantum circuit recompilation can be viewed as a \new{method of reducing noise}. Furthermore, if we recompile a circuit without knowledge of individual gate times, we can reformulate the problem generally as finding a circuit which implements $\hat{U}$ in  $o < m$ gate operations. Solutions to this range from duplicate gate cancellation \cite{miller2010lowering} to two-qubit block re-synthesis \cite{jurcevic2021demonstration} involving the KAK decomposition \cite{bullock2003arbitrary}.

In this work we utilise recent advances in approximate circuit recompilation. Here, rather than find an exact alternative implementation of $\hat{U}$, the goal is to find a shallower quantum circuit $\hat{V}$ which has approximately the same action on some initial state $\hat{U}|\psi\rangle \approx \hat{V}|\psi\rangle$. Reformulating this as $\langle\psi|\hat{V}^\dagger \hat{U}|\psi\rangle \approx 1$, we notice that the problem can be viewed as finding the set of gates $\hat{V^\dagger}$ that inverses the action of $\hat{U}$ as measured by the overlap between the initial and final state. If the initial state is $|0\rangle^{\bigotimes n}$, as is convention for quantum algorithms, then the problem can be solved with incremental structural learning (ISL) \cite{jaderberg2020minimum}. In ISL, the structure of $\hat{V^\dagger}$ is informed by incrementally adding gates that work to disentangle the original circuit back to the $|0\rangle^{\bigotimes n}$ state, leading to significant gate count reductions \cite{jaderberg2021quantum} particularly for Trotterised time evolution circuits \cite{fitzpatrick2021evaluating}. More details of ISL can be found in \cite{jaderberg2020minimum}. Despite being a quantum algorithm, in this work ISL can be viewed as a classical method, since we sample the state $\hat{U}|0\rangle^{\bigotimes n}$ on classically simulated quantum hardware for better performance. Overall, ISL can be considered as a method for substituting circuit depth for increased circuit evaluations, such that obtaining the population at a given time requires $O(kn_e)$ rather than $O(k)$ evaluations, where $k$ is the number of shots and $n_e$ is the number of evaluations required for convergence of ISL. Analysis of how the value of $n_e$ changes for different systems is an ongoing effort~\cite{fitzpatrick2021evaluating}, \new{yet ultimately its scaling is unknown, a limitation shared with other variational quantum algorithms including the variational quantum eigensolver \cite{tilly2021variational}. Numerically,} across this work the value of $n_e$ ranges from $\sim10^1$ to $\sim10^4$ depending on the complexity and depth of the circuit.

\begin{figure}
	\centering
	\includegraphics[width=15cm,valign=b]{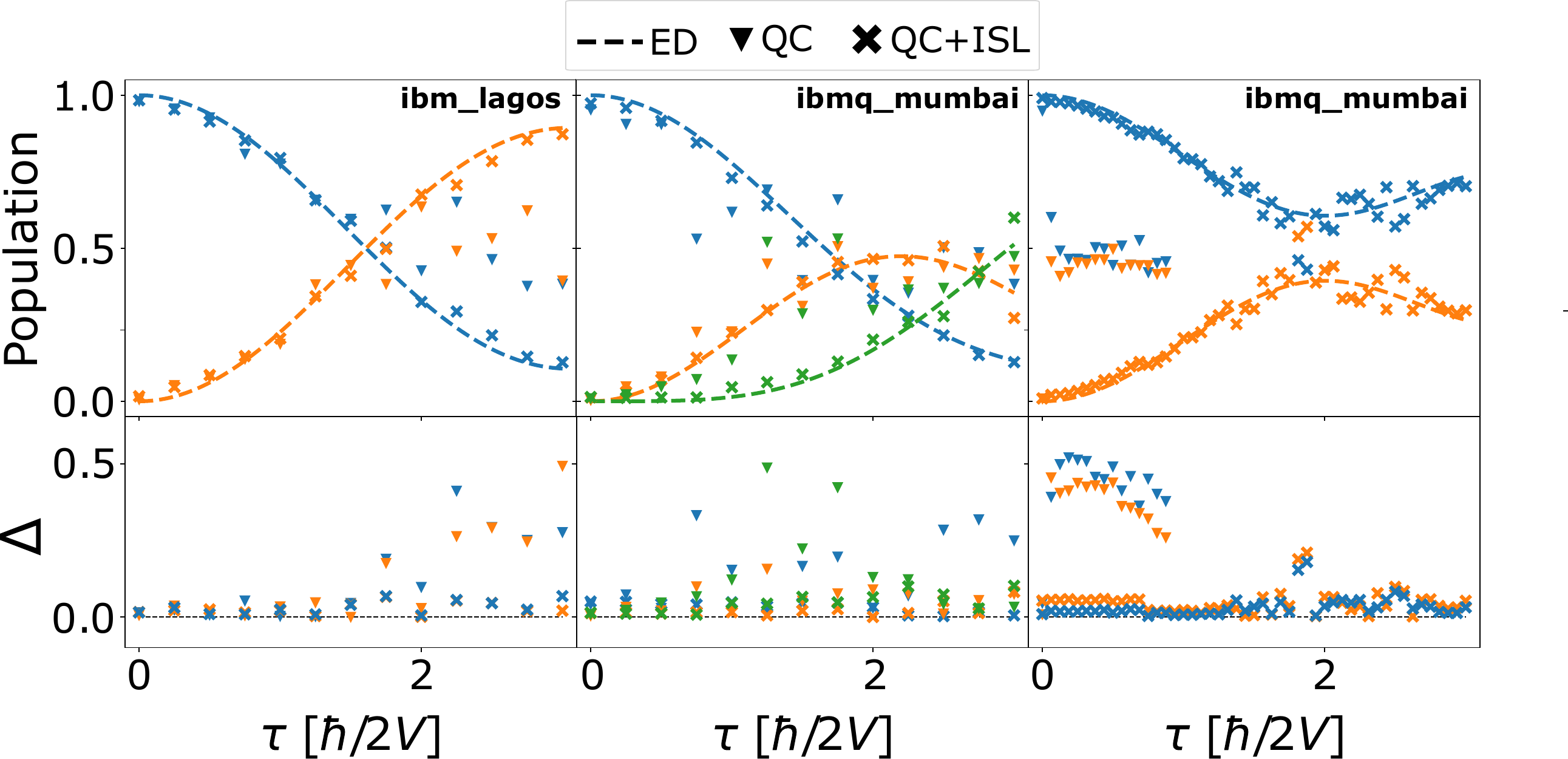}
	\caption{\new{Population} over time as calculated by ED (dashed line) or quantum circuits executed on an IBM quantum computer directly (triangle marker) or first recompiled with ISL (crossed marker). All experiments are in the strong electronic coupling regime $V=1.0\hbar\omega$. Left: system parameters $N=2, \chi=0.3\hbar\omega, \eta=12$ and $n_x=1$ run on \textit{ibm\_lagos}. Middle: same as the left panel, but with $N=3$ electronic sites, run on \textit{ibmq\_mumbai}. Right: $N=2, \chi=1.0\hbar\omega, \eta=48$ and $n_x=2$ run on \textit{ibmq\_mumbai}. \new{The bottom panel shows the absolute difference $\Delta$ between experiments and the analytic solution for each site.}
	\label{fig:v1_realdevice}}
\end{figure}

To see what simulations are achievable on a real quantum computer, we first try the regime from section \ref{subsec:ED_comparison_results} which achieves convergence with the fewest quantum resources. For this we consider a two-site system with couplings $V=1.0\hbar\omega$, $\chi=0.3\hbar\omega$, which produces accurate results with as few as $n_x=1$ qubit per oscillator (4 qubits total) and $\eta=12$ Trotter steps. The left panel of Figure \ref{fig:v1_realdevice} shows the \new{population}, obtained through experiments on the 7 qubit, 32 QV (quantum volume) \textit{ibm\_lagos} quantum computer with $k=8192$ shots per circuit. \new{For all the experiments on real hardware, a measurement calibration scheme was performed to correct for readout errors.} The triangle markers indicate the results obtained when implementing the Trotterised time evolution operator without approximate circuit recompilation. Due to the small size of the system, only 6 CNOT and 12 single-qubit gates are required per Trotter step, producing accurate results for up to 6 Trotter steps of evolution. For this simple problem, that corresponds to half the evolved time due to its low total Trotter step requirements. Yet the number of gate operations still grows linearly with the number of Trotter steps, such that simulating 7 through 12 Trotter steps of evolution, requiring 42 to 72 CNOT gates respectively, leads to a sharp decrease in accuracy. Here we focus on the number of two-qubit gates, as opposed to single-qubit gates, because their error rate is an order of magnitude higher on the devices used.

By contrast, the crossed markers show the results obtained running on the same device but first recompiling the circuits with ISL. Here, for \textit{any} number of Trotter steps, ISL finds an equivalent circuit producing the same state to 99$\%$ overlap with on average 3.4 CNOT gates and 8.8 single-qubit gates. \new{An example walk-through of the steps ISL takes to reach this solution for one of the time evolution circuits can be found in \ref{app:isl_stepwise}. Here we give an average number of operations across all the 12 different time evolution circuits, since the use of ISL decouples the scaling of the number of gates with Trotter steps from the previous linear relationship. Instead, in many cases the ISL solution can have constant or even reduced depth when adding Trotter steps and it is not obvious that later-time states should always be more difficult to recompile. A good example of this would be the recompilation of periodic systems, which may return at later time to less entangled states.} Excitingly, the low average number of operations produced by the ISL solution enables the results obtained on the real device to accurately reproduce those calculated via exact diagonalisation, even in the region $\tau > 1.5$.

Following this, we simulate a larger system to further test what is possible on a real quantum computer. We repeat the previous simulation but with $N=3$ electronic sites, giving a total of 6 qubits required. Whilst the inclusion of extra qubits itself incurs no error, the additional gates required to simulate a larger chain and another oscillator mode leads to more noise. The middle panel of Figure \ref{fig:v1_realdevice} shows results obtained on the 27 qubit, 128 QV \textit{ibmq\_mumbai} device with $k=8192$ shots per circuit. Here, a direct implementation of the time evolution requires 10 CNOT gates and 20 single-qubit gates per Trotter step, the results of which are shown by the triangle marker. Despite running on a higher fidelity quantum computer, the increased number of gates per Trotter step are such that the \new{population} obtained at each site is only accurate for the first few Trotter steps. Beyond this the accuracy decreases, such that for the 12 Trotter step circuit used to calculate the population at $\tau=3.0$, requiring 120 CNOT gates, the qubit measurement statistics are random. We note that a random qubit state translates to a measured population of 0.5, which in this graph is deceivingly close to the real dynamics at later time steps.

When recompiling this larger system, ISL finds for any number of Trotter steps an equivalent circuit producing the same state to 95$\%$ overlap with on average 10.0 CNOT gates and 18.7 single-qubit gates. Although this is more operations than the two site case, it is sufficiently shallow so that the noisy device can produce dynamics similar to that of ED for all but two of the observed points. The increased error at $\tau=2.5$ and $3.0$ can be explained by the observation that the recompiled solutions at these times contained more operations than average, with 21 and 16 CNOT gates respectively.

As mentioned in section \ref{subsec:ED_comparison_results}, although the low Trotter step and $n_x$ requirements of the small $\chi$ regime make it appealing for quantum simulation, it is unlikely one would require non-classical methods for such simple dynamics. Therefore, in this final experiment we test what can be simulated on a real quantum computer for the nontrivial strongly coupled case $V=1.0\hbar\omega, \chi=1.0\hbar\omega$. Here we use $\eta=48$ Trotter steps as required for convergence and $n_x=2$, which suffers from very minimal truncation at the benefit of less qubits and gates. In this case the limitation of the direct implementation becomes apparent, with each Trotter step requiring 42 CNOT and 66 single-qubit gates. Furthermore, because this regime requires $n_x>1$ qubits per oscillator for convergence, the graph of connected qubits in the circuit no longer matches the honeycomb structure of the hardware used in this work. Mapping between the topologies requires SWAP operations which dramatically increases the number of CNOT gates per Trotter step to 133. The result of executing these circuits on \textit{ibmq\_mumbai} is shown by the triangle markers in the right panel of Figure \ref{fig:v1_realdevice}. Here, it is not possible to obtain an accurate population after even 1 Trotter step of evolution, and the signal rapidly decays to random noise. Conversely, the populations obtained using ISL match the ED results well up to $\tau=1.5$ and follow the correct trend after. This is possible because the recompiled circuits allow us to obtain the dynamics for any point in time up to $\tau=3.0$ with on average 15.6 CNOT gates and 26.9 single-qubit gates. Furthermore, the use of SWAP gates can be avoided by restricting the qubit connections allowed during ISL to only those which are connected on the physical device.

\section{Conclusion} \label{sec:conclusion}

In this work we demonstrate the first simulation of electron-phonon dynamics on a real quantum computer. To achieve this, we first establish the previously unknown Trotter step and qubit requirements for accurate simulation of systems described by the electron-phonon Hamiltonian, which we find grows with the ratio of the electron-phonon to dipole-dipole couplings $\chi/V$. We then run the converged time evolution circuits on real quantum computers and demonstrate that, despite the resource efficiency of the scheme, accurate results cannot be obtained even for small systems. To remedy this, we use incremental structural learning to find approximately equivalent circuits for each time step in the simulation, leading to a significant reduction in the number of gates required. Executing the recompiled circuits on a real quantum computer, we obtain highly accurate dynamics for small systems with weak electron-phonon coupling and partially accurate results for the more complicated case of strong coupling.

Evolving electron-phonon systems on a quantum computer is not only beneficial for the direct study of dynamics, but also for optimisation routines whose cost functions contain time dependent observables of such systems. Important example of this include studying the conditions required for optimal energy transfer \cite{rebentrost2009environment} and fitting values to unknown system parameters based on experimental data \cite{knoester1993nonlinear}. Notably, solving such problems can be done in the presence of noise \cite{fontana2021evaluating,moll2018quantum}, making it an appealing application for near-term quantum computers. Thus, using real quantum hardware to solve optimisation problems involving electron-phonon systems would represent an interesting extension of the results presented in this work.

In the long term, the study of electron-phonon dynamics will benefit greatly from simulation on error-corrected quantum computers and its associated exponential speedup. However, the ongoing development of quantum hardware demonstrates that the ability to execute deep circuits will remain a longstanding problem to obtain accurate results for interesting physical systems. In this way, our work provides a significant first step to address this, and our novel approach illustrates that electron-phonon Hamiltonians can be simulated on near-term quantum computers when using approximate circuit recompilation. Looking forward, the question remains as to whether sufficiently large systems could be simulated as to achieve a quantum advantage using our method. This would require the execution of the ISL routine itself on quantum hardware, which in turn involves evaluating quantum circuits with depth beyond the capabilities of current devices. Nevertheless, the rapid rate of hardware progress may soon change this, and as such attempting to scale the methods put forward in this paper is an imperative and exciting avenue of future work.

\section{Data availability}
Data used to generate the above figures are available upon request from the authors.

%\begin{acknowledgments}
\ack
	B.J. and D.J. acknowledge support from the EPSRC National Quantum Technology Hub in Networked Quantum Information Technology (EP/M013243/1) and the EPSRC Hub in Quantum Computing and Simulation (EP/T001062/1).
%\end{acknowledgments}

\appendix

\section{Conversion of qubit operators to Pauli matrices}\label{app:qubit_to_pauli}

Given a set of operators acting on qubit computational basis states, they can be mapped to Pauli matrices via the relations

\begin{subequations}\label{eqn:binary_to_pauli}
    \begin{align}
        |0\rangle \langle0| &=  \frac{1}{2}\left(\mathbbm{1} + \hat{\sigma}^z \right)\\
        |0\rangle \langle1| &=  \frac{1}{2}\left(\hat{\sigma}^x + \imag\hat{\sigma}^y \right)\\
        |1\rangle \langle0| &=  \frac{1}{2}\left(\hat{\sigma}^x - \imag\hat{\sigma}^y \right)\\
        |1\rangle \langle1| &=  \frac{1}{2}\left(\mathbbm{1} - \hat{\sigma}^z \right).
    \end{align}
\end{subequations}

\section{Conversion of ladder operators to Qubit operators}\label{app:ladder_to_qubit}

Here we demonstrate how to express the ladder operators of the QHO in terms of qubit operators. For example, the term $\hat{a}^\dagger + \hat{a}$ in Eq. (\ref{eqn:Hep}) can be expanded as

\begin{equation}\label{eqn:expanding_adagger_plus_a}
    \hat{a}^\dagger + \hat{a} = \sum_{d=0}^{2^{n_{x}}-2}\sqrt{(d+1)}|d+1\rangle \langle d| + \sum_{d=1}^{2^{n_{x}}-1}\sqrt{d}|d-1\rangle \langle d|,
\end{equation}

where $|d\rangle$ is the eigenstate of the $d^{th}$ excited state. Considering the specific example of $n_x=2$ $(d=4)$, we obtain

\begin{equation}\label{eqn:x_operator_binary}
    \begin{aligned}
        \hat{a}^\dagger + \hat{a} = & \; |01\rangle \langle 00| + |00\rangle \langle 01|\\& + \sqrt{2}(|10\rangle \langle 01| 
         + |01\rangle \langle 10|)\\& + \sqrt{3}(|11\rangle \langle 10| + |10\rangle \langle 11| ).
    \end{aligned}
\end{equation}

Subsequently, the action on each qubit can be mapped to Pauli operators as shown in \ref{app:qubit_to_pauli}. As an example in the context of ladder operators, the first two terms of Eq. (\ref{eqn:x_operator_binary}) would then become

\begin{equation}\label{eqn:pauli_string_example}
   |01\rangle \langle 00| + |00\rangle \langle 01| = \frac{1}{2}\left(\hat{\sigma}^x_1 + \hat{\sigma}^z_2\hat{\sigma}^x_1\right)
\end{equation}

after cancellations.

\section{Population with no phonon coupling}\label{app:chi_0_results}

\begin{figure}
	\centering
	\includegraphics[width=8cm]{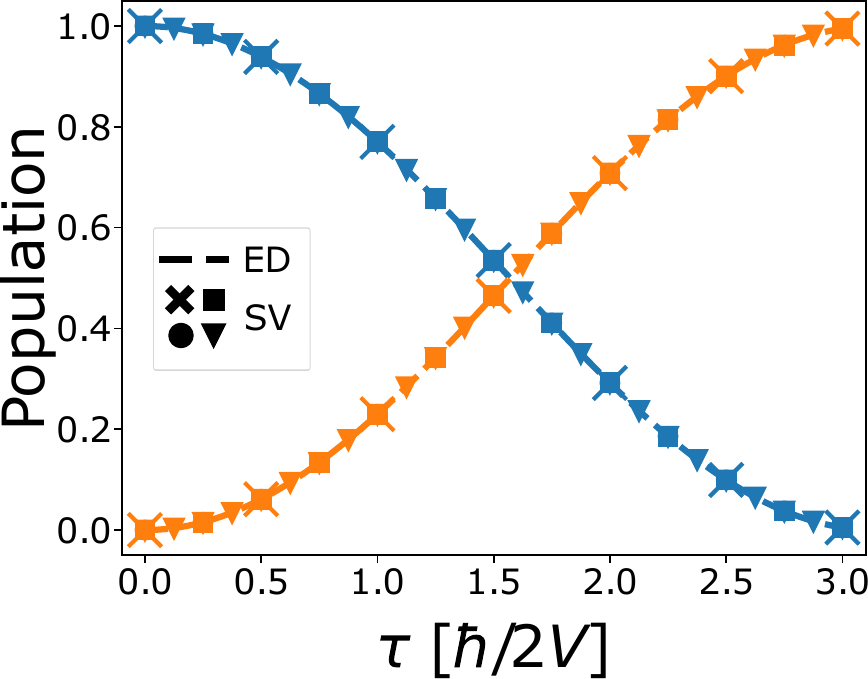}
	\caption{Population over time at electronic sites 0 (blue) and 1 (orange), as calculated by ED (dashed line) or quantum circuits run on the Qiskit Aer statevector simulator (markers) with 6 (cross), 12 (square), 24 (circle) or 48 (triangle) Trotter steps. The Hamiltonian is simulated with parameters $V=0.05\hbar\omega$, $\chi=0.0\hbar\omega$. In this case there is no coupling to the phonon modes, so the system evolves under just $H = H_{\mathrm{el}}$. Since there are no non-commuting terms, the quantum circuit method incurs zero Trotter error and thus achieves equal accuracy to ED for any number of Trotter steps. Note that consequentially the dashed line is partially obscured by the symbols.}
	\label{fig:x=0}
\end{figure}

Figure \ref{fig:x=0} shows the site \new{populations} over time for the case of weak electronic coupling $V=0.05\hbar\omega$
and no phonon coupling. Here, as expected for the purely electronic case, the exact diagonalisation results (dashed line) demonstrate a simple pattern of the excitation transferring from its initial placement on site 0 to site 1. The different markers represent obtaining the population using the quantum circuit method described in the main text, repeated with increasing numbers of Trotter steps. Here the lack of electron-phonon coupling simplifies things, since the system effectively evolves under the Hamiltonian $H=H_{el}$. Therefore, without any non-commuting terms in the Hamiltonian, the Trotter error is zero and adding more Trotter steps does not improve the accuracy of the solution with respect to the ED results.

\section{Simulator results of larger real world system}\label{app:large_results}

In the main text, we presented results using a statevector simulator for systems with two electronic sites, one phonon mode per site and up to $n_x=4$ for a maximum of ten qubits. This allowed us to study different coupling regimes and by keeping the Hilbert space small, compare them easily to exact diagonalisation results. In this section however, we demonstrate the flexibility of the quantum circuit approach to scale to much larger systems. Unlike analogue simulation of the electron-phonon Hamiltonian, our method requires no redesigning of the algorithm or hardware to scale up, since the circuits are created using the general rules laid out in section \ref{subsec:circuit_implementation}. Furthermore, whilst our previous simulations had uniform couplings, here we can consider systems with non-uniform couplings with no additional cost.

\begin{figure}
	\centering
	\includegraphics[width=8cm,valign=b]{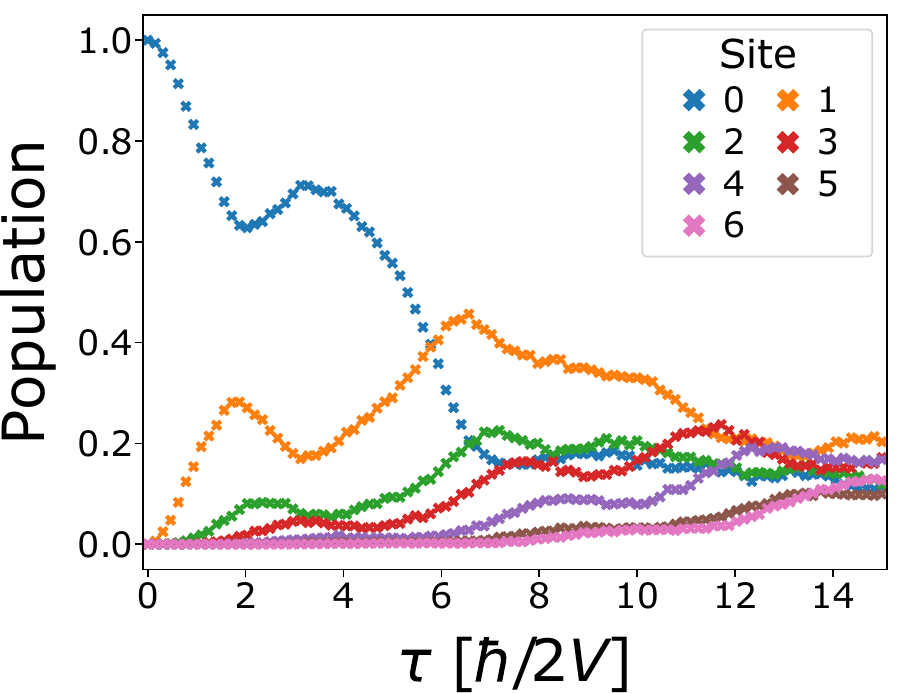}   
	\caption{Population over time with non-uniform couplings and phonon frequencies, $\eta=96$ and $n_x=3$. All results were calculated using the quantum circuit implementation described in the main body of this work, run on the Qiskit qasm simulator with 10,000 shots. The couplings $\vec{V}$, $\vec{\chi}$ and frequencies were randomly generated from the uniform interval [0.8, 1.2] and can be found in \ref{app:random_params_08_12}.
	\label{fig:random_nx3_n7_08_12}}
\end{figure}

Figure \ref{fig:random_nx3_n7_08_12} shows the result of our larger scale simulation, run on the qasm simulator with $k=10,000$ shots. Here we have $N=7$ electronic sites, $l=1$ phonon mode per site and $n_x=3$ qubits per oscillator for a total of 28 qubits. The couplings $\vec{V}$ and $\vec{\chi}$ are randomly generated from a uniform distribution in the range $[0.8, 1.2]\hbar\omega$ and can be found in \ref{app:random_params_08_12}. We choose this range such that the mean value still represents the strong coupling regime, which we found in section \ref{subsec:ED_comparison_results} to have interesting and non-trivial dynamics. 

Looking at Figure \ref{fig:random_nx3_n7_08_12} up to $\tau=3$, we see a pattern whereby the excitation originally localised to site 0 distributes down the chain. This is the primary excitation wave, causing visible peaks on sites 1, 2 and 3 on this short time scale. However, beyond this point the dynamics become much more complex, with many oscillations and peaks of different scales. This is caused first by the reflection of the primary wave off each site, and subsequently by the complex interference between counterpropogating wavepackets. By the end of our simulation, the net transfer across the sites begins to slow, such that the dynamics appear to begin reaching an equilibrium.

\section{Figure \ref{fig:random_nx3_n7_08_12} parameters}\label{app:random_params_08_12}

Here we present the random couplings used to generate Fig. \ref{fig:random_nx3_n7_08_12}. Note that the structure of the $\vec{V}$ matrix corresponds to electronic sites connected in a chain topology. One could translate this to a ring topology by adding non-zero values in the lower-left and upper-right corners of the matrix, or a fully connected topology by filling all elements with non-zero values.

\begin{equation}
\vec{V}=
\begin{bmatrix}{}
0 & 0.981 & 0 & 0 & 0 & 0 & 0 \\
0.981 & 0 & 1.024 & 0 & 0 & 0 & 0 \\
0 & 1.024 & 0 & 1.170 & 0 & 0 & 0 \\
0 & 0 & 1.170 & 0 & 0.986 & 0 & 0 \\
0 & 0 & 0 & 0.986 & 0 & 1.003 & 0 \\
0 & 0 & 0 & 0 & 1.003 & 0 & 1.035\\
0 & 0 & 0 & 0 & 0 & 1.035 & 0 
\end{bmatrix}\hbar\omega
\end{equation}
\begin{equation}
\vec{\chi}=
\begin{bmatrix}{}
1.149\\
1.002\\
1.192\\
1.079\\
1.089\\
0.963\\
1.024\\
\end{bmatrix}\hbar\omega
\end{equation}

\new{\section{Stepping through ISL recompilation}\label{app:isl_stepwise}}

\new{In the main text, approximate circuit recompilation is a crucial tool that allows the substitution of time-evolution circuits with significantly shallower alternatives. Whilst the original scheme is defined in \cite{jaderberg2020minimum}, here we provide an explicit example of the steps of the ISL algorithm when solving a target circuit that generates the dynamics studied in the left panel of Fig. \ref{fig:v1_realdevice}. In this particular model we have system parameters $N=2, \chi=0.3\hbar\omega, n_x=1$ and the overall time is divided up into $\eta=12$ Trotter steps. Given we wish to obtain the population at each time step, our simulation algorithm in the main text requires evaluating 13 different circuits each with a different integer $\zeta < \eta$ repetitions of the Trotter step circuit before measurement. For the demonstration of ISL, we will consider a recompilation target circuit evaluating the dynamics at $\zeta=4$ time slices into the evolution.}

\begin{figure}
	\centering
	\centerline{\includegraphics[width=1.1\linewidth]{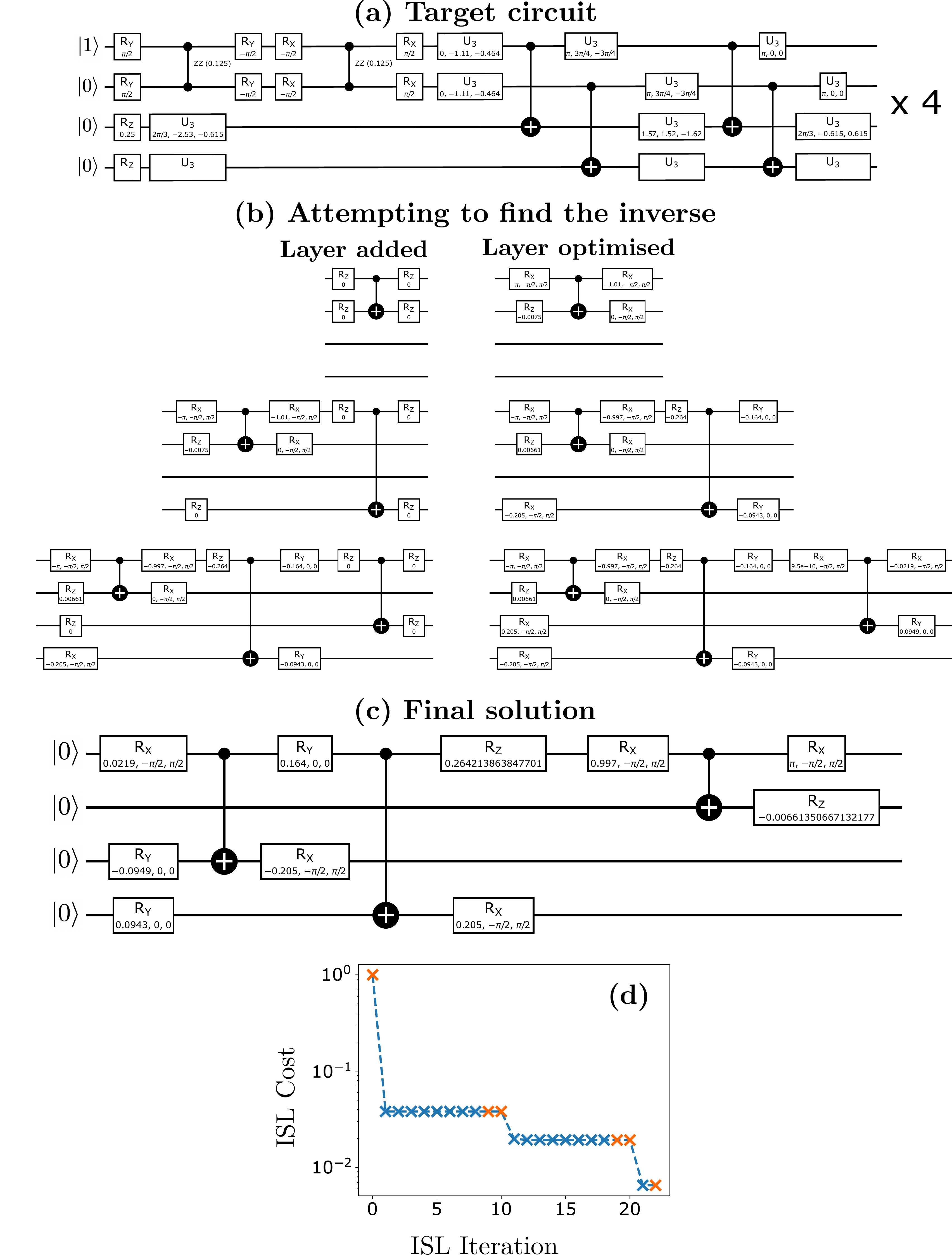}}
	\caption{Caption next page.}
	 \label{fig:isl_stepwise_solving}
\end{figure}

\new{Fig. \ref{fig:isl_stepwise_solving} demonstrates the overall process. At the top of the figure the target circuit to recompile is defined. This matches the structure of the general time evolution circuit laid out in Fig. \ref{fig:ts_circuit}, but with higher-level gates decomposed into standard quantum computing operations with explicit rotation angles including the single-qubit unitaries $\hat{U}_{1}$, $\hat{U}_{2}$ and $\hat{U}_{3}$ defined as~\cite{qiskit}}

\new{\begin{align}
\hat{U}_{3}(\theta, \phi, \lambda) & = \begin{pmatrix} \cos(\theta/2) & 
-e^{i\lambda}\sin(\theta/2) \\ e^{i\phi}\sin(\theta/2) & 
e^{i\lambda+i\phi}\cos(\theta/2) \end{pmatrix}\,, \\
\hat{U}_{2}(\phi, \lambda) & = \hat{U}_{3}(\pi/2, \phi, \lambda) = 
\frac{1}{\sqrt{2}} \begin{pmatrix} 1 & -e^{i\lambda} \\ e^{i\phi} & e^{i(\phi + 
	\lambda)} \end{pmatrix}\,,\\
\hat{U}_{1}(\lambda) & = \hat{U}_{3}(0, 0, \lambda) = \begin{pmatrix} 1 & 0 \\ 
0 & e^{i \lambda} \end{pmatrix}\,.
\end{align}}

\new{The circuit diagram at the top of Fig. \ref{fig:isl_stepwise_solving} shows the gates required for one Trotter step. In this example we will choose our target circuit to be one that simulates the population after four Trotter steps of evolution and as such we repeat the circuit four times. Additionally, we note the first qubit starts in the $|1\rangle$ state, reflecting the physical system under study from section \ref{subsec:real_device_results} with an electronic excitation localised on the first site. Since ISL assumes the initial state $|\psi_0\rangle = |0\rangle^{\bigotimes n}$, this state preparation is incorporated into the recompiled solution. This means that recompilation would need to be repeated if studying systems with different initial states.}

\new{Moving downwards across the figure, the ISL procedure begins and layers of gates are incrementally appended to the target circuit with the goal of inversing the action of the time evolution circuit. Each layer consists of a CNOT gate surrounded by 4 single-qubit rotation gates with parameterised rotation axis and angles, initialised as $\hat{R}_z(0)$. First, a layer is added between the two qubits with the highest pairwise entanglement, for which we use negativity \cite{vidal2002computable} as our measure. Subsequently the axes and angles of rotation of the layer are optimised using the Rotoselect algorithm \cite{ostaszewski2021structure} with respect to minimising the cost}

\new{\begin{equation}\label{eqn:isl_cost}
   C = 1 - \left|\bra{\psi_{0}}\hat{V}^{\dagger}\hat{U}\ket{\psi_{0}}\right|^2  
\end{equation}}

\new{where $\hat{U}$ is the target circuit, $\hat{V}^{\dagger}$ is the current best guess of the inverse and $|\psi_0\rangle = |0\rangle^{\bigotimes n}$ is the initial state of the the circuit. Once optimising the layer has converged the cost function is evaluated once again. If the cost is now below the threshold $C<C_{t}$, this part of the algorithm is completed. If not, then another layer is added and the optimisation procedure is repeated for the new layer. In this particular example, it takes three layers before the cost threshold $C_t = 0.01$ is reached. At this point, the ISL algorithm is terminated and the circuit that represents the best guess of $\hat{V}^{\dagger}$ is recursively inverted gate-by-gate to produce the final recompilation solution $\hat{V}$. Some simple non-approximate techniques are also applied to improve the final solution, including the merging of adjacent rotation gates in the same basis, removing two-qubit gate blocks that form an identity (e.g., consecutive CNOT gates) and removing of any gates for which the rotation angle is below 0.001.}

\new{The bottom of the figure shows the value of the cost at each iteration in the ISL procedure, with the orange markers corresponding to the visualised circuits above. The process starts with the first orange marker, corresponding to the first layer added with initial rotations $\hat{R}_z(0)$. Next, a series of 8 blue markers, corresponding to iterations of the Rotoselect algorithm, sees an immediate drop in the cost followed by several steps of small refinement. The convergence of the first layer is then given by the second orange marker, which corresponds to the second circuit diagram of the trial solutions. This process of adding a layer, a large jump on the first optimiser iteration and then subsequent refinement is repeated two more times until the cost threshold of 0.01 is met.}

\new{In this Appendix we have stepped through recompilation of the circuit corresponding to four Trotter steps of evolution of the electron-phonon system studied in the main text with system parameters $N=2, \chi=0.3\hbar\omega, n_x=1$ and $\eta=12$. Evaluating the target circuit and the recompiled solution gives the population at just one point in time corresponding to the fourth triangle and crossed marker of Fig. \ref{fig:v1_realdevice} respectively. To obtain the population at each point and produce the whole figure, recompilation needs to be repeated for 0 through 12 Trotter steps of evolution. We note that there are two ways to achieve this. The first is the simplest and involves recompilation from scratch for every point in time. The second is so called ladder-ISL, in which each Trotter step in the circuit is sequentially recompiled, with the recompiled solution to the previous Trotter step used as an approximation of the evolution up to the next Trotter step. More details of this approach can be found in \cite{jaderberg2020minimum}. Whilst ladder-ISL is more amenable to near-term quantum computers, since the evaluated circuits are shallower,  this approach introduces an additional error term, since the approximation at each Trotter step accumulates. Therefore, in this proof-of-principle work on smaller systems, we choose to recompile each time step from scratch for maximum accuracy.}

\addtocounter{figure}{-1}
\begin{figure} [t!]
  \caption{A step-by-step walk-through of the usage of ISL in this work. (a) In this example, the target circuit corresponds to four Trotter steps of evolution of the electron-phonon Hamiltonian with system paramters $N=2, \chi=0.3\hbar\omega, \eta=12$ and $n_x=1$. (b) The ISL procedure begins by appending a layer of gates to the target circuit that is as a trial solution of the inverse $\hat{V}^{\dagger}$. The rotation axes and angles of this layer are then then optimised with respect to producing a final state which maximally overlaps the input state as given in Eq. (\ref{eqn:isl_cost}). After 3 layers the cost is below the threshold value of 0.01 and this process is finished. (c) The circuit at termination of the previous step is then recursively inverted to produce the solution to the recompliation problem $\hat{V}$. (d) The value of the cost function at each ISL iteration, defined as either an optimisation iteration or adding a layer. The crossed markers in orange correspond to the cost evaluated for the circuits visualised in (b).}
\end{figure}

\newpage
\section*{References}
\bibliographystyle{journal_v5.bst}
\bibliography{bibliography.bib}

\end{document}